\documentclass[final,5p,times,twocolumn]{elsarticle}

\usepackage{graphicx}

\usepackage{amssymb}

\journal{Journal of Physics and Chemistry of Solids}

\begin{document}

\begin{frontmatter}



\title{Upper critical field of the 122-type iron pnictide superconductors}

\author[a]{L. Jiao}
\author[a]{J.L. Zhang}
\author[b]{F.F. Balakirev}
\author[c]{G.F. Chen}
\author[c]{J.L. Luo}
\author[c]{N.L. Wang}
\author[a]{H.Q. Yuan\corref{cor1}}

\address[a]{Department of Physics, Zhejiang University, Zhejiang 310027, P. R. China}
\address[b]{NHMFL, Los Alamos National Laboratory, MS E536, Los Alamos, NM 87545, USA}
\address[c]{Beijing National Laboratory for Condensed Matter Physics, Institute of Physics, Chinese Academy of Science, Beijing, 100080, P. R. China}
\ead{hqyuan@zju.edu.cn}
 \cortext[cor1]{Corresponding author.}

\begin{abstract}
The upper critical fields ($H_{c2}$) of the single crystals
$\rm(Sr,Na)Fe_2As_2$ and $\rm Ba_{0.55}K_{0.45}Fe_2As_2$ were
determined by means of measuring the electrical resistivity, $
\rho_{xx}(\mu_0H)$, using the facilities of pulsed magnetic field at
Los Alamos. In general, these compounds possess a very large upper
critical field ($H_{c2}(0)$) with a weak anisotropic effect. The
detailed curvature of $H_{c2}(T_c)$ may depend on the magnetic field
orientation and the sample compositions. We argue that such a
difference mainly results from the multi-band effect, which might be
modified via doping.
\end{abstract}


\end{frontmatter}



\section{Introduction}
\label{} The discovery of high temperature superconductivity in $\rm
LaO_{1-x}F_xFeAs (\emph{x}=0.05-0.12)$~\cite{KamiharaWatanabe} has
stimulated considerable research efforts on the search of new
superconducting materials and the elucidation of its pairing
mechanism. Resembling the copper oxides, the iron pnictides
crystalize in a layered structure and superconductivity occurs upon
suppressing the magnetic order either by chemical
doping~\cite{Rotter38,ChenXH} or by applying
pressure~\cite{Park,Alireza}. However, these new superconductors
also exhibit their own unique properties. Instead of a
Mott-insulator as observed in cuprates, the parent compounds of iron
pnictides are usually bad metals~\cite{Si}. Furthermore, nearly
isotropic superconductivity~\cite{Yuan565,Fang} and three
dimensional energy dispersion~\cite{ding} have been observed in some
iron-based superconductors even though they possess a layered
crystal structure. These findings are surprising and deserve further
research in order to check their generality.

In a multi-band system, the electronic properties might strongly
depend on the electron/hole doping level, which would accordingly
change the anisotropy of the upper critical field. Experimental
confirmation on it might further support the multi-band
superconductivity in iron pnictides. Moreover, the value of $H_{c2}$
can be largely enhanced by introducing disorder in a multi-band
superconductor, e.g. in $\rm MgB_2$~\cite{Gurevich}. One related
question is whether the large $H_{c2}(0)$ observed in iron pnictides
is intrinsic or an effect of disorder. Elucidation of these
questions would help to catch the intrinsic information of
superconductivity in iron pnictides.

Here we report the measurements of magnetoresistance in single
crystals of $\rm(Sr,Na)Fe_2As_2$ and $\rm
Ba_{0.55}K_{0.45}Fe_2As_2$. It is indeed found that the upper
critical field of all these compounds show rather weak anisotropic
effect. The observation of huge $H_{c2}(0)$ ($\sim$130T) in $\rm
Ba_{0.55}K_{0.45}Fe_2As_2$, whose sample quality has been improved
in comparison with those previously presented in Ref~\cite{Yuan565},
indicates that the large upper critical field is an intrinsic and
general feature of iron pnictides.

\section{Experimental Methods}
Single crystal of $\rm(Sr,Na)Fe_2As_2$ and $\rm
Ba_{0.55}K_{0.45}Fe_2As_2$ were synthesized by solid state reaction
method using FeAs as flux~\cite{ChenGF78}. The derived crystals were
characterized to be a single phase by powder x-ray diffraction (XRD)
with Cu $K\alpha$ radiation at room temperature. The doping
concentration of K in $\rm Ba_{0.55}K_{0.45}Fe_2As_2$ is nominal
value.

The electrical resistance at zero field, $R(T)$, was measured with a
Lakeshore ac resistance bridge. The magnetic field dependence of the
resistivity, $R_{xx}(H)$, was measured up to 60T using a typical
4-probe method in a capacitor-bank-driven pulsed magnet. The data
traces were recorded on a digitizer using a custom designed
high-resolution, low-noise synchronous lock-in technique. In order
to minimize the inductive heating caused by a pulsed magnetic field,
small crystals with typical size 2mm$\times$0.5mm$\times$0.1mm were
cleaved off along the \emph{c}-direction from the as-grown samples.

\section{Results and Discussion}
\begin{figure}
 \includegraphics[width=0.45\textwidth]{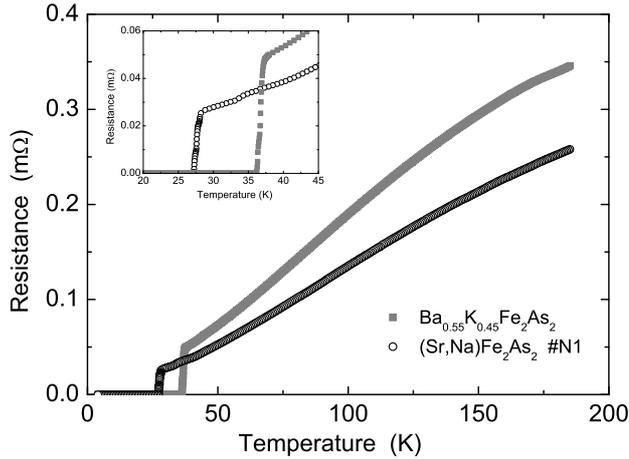}
\caption{Temperature dependence of the electrical resistance for
$\rm(Sr,Na)Fe_2As_2$ (\#N1) and $\rm Ba_{0.55}K_{0.45}Fe_2As_2$,
respectively. The insert enlarges the section at the superconducting
transition.} \label{fig1}
\end{figure}

Fig.\ref{fig1} shows the electrical resistance at zero field for
$\rm(Sr,Na)Fe_2As_2$(\#N1) and $\rm Ba_{0.55}K_{0.45}Fe_2As_2$,
respectively. It is noted that no anomalies of structural/magnetic
phase transitions are visible at high temperature, suggesting that
these samples are likely located in the optimal doing region.The
superconducting transition temperature, determined from the
mid-point of the superconducting transition, gives $T_c=27.6$K and
36.8K for $\rm(Sr,Na)Fe_2As_2$(\#N1) and $\rm
Ba_{0.55}K_{0.45}Fe_2As_2$, respectively. The sharp superconducting
transitions (see the inset of Fig.\ref{fig1}) indicate a good sample
quality for both $\rm(Sr,Na)Fe_2As_2$(\#N1) and $\rm
Ba_{0.55}K_{0.45}Fe_2As_2$. In comparison with those samples
reported in Ref~\cite{Yuan565}, $\rm Ba_{0.55}K_{0.45}Fe_2As_2$
studied here shows a higher $T_c$ and a larger value of residual
resistivity ratio (\emph{RRR}), further supporting the improvement
of sample quality.

\begin{figure}\centering
\includegraphics[width=0.45\textwidth]{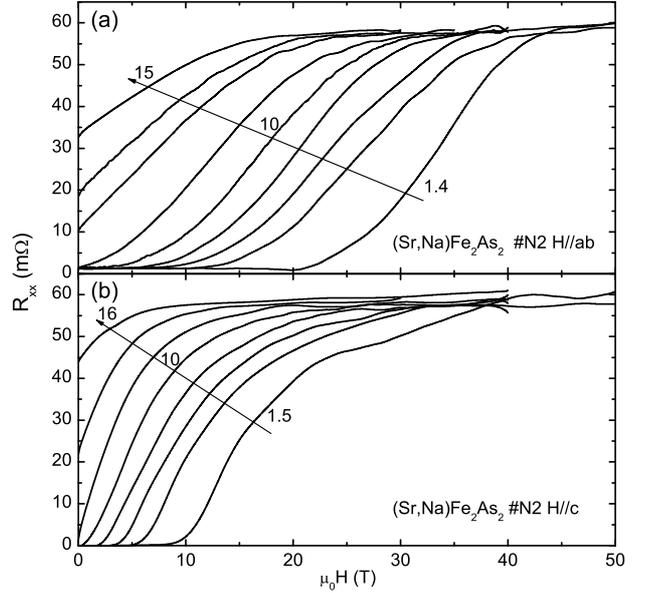}
\caption{Magnetic field dependence of the electrical resistance at
variant temperatures for $\rm (Sr,Na)Fe_2As_2$ (\#N2). (a) $H//ab$
and $T$=1.4K, 4.7K, 6K, 8K, 10K, 12K, 13K, 14K, 15K; (b) $H//c$ and
$T$=1.5K, 4K, 6K, 8K, 10K, 12K, 14K, 16K (from bottom to top).}
\label{fig2}
\end{figure}

\begin{figure}\centering
\includegraphics[width=0.45\textwidth]{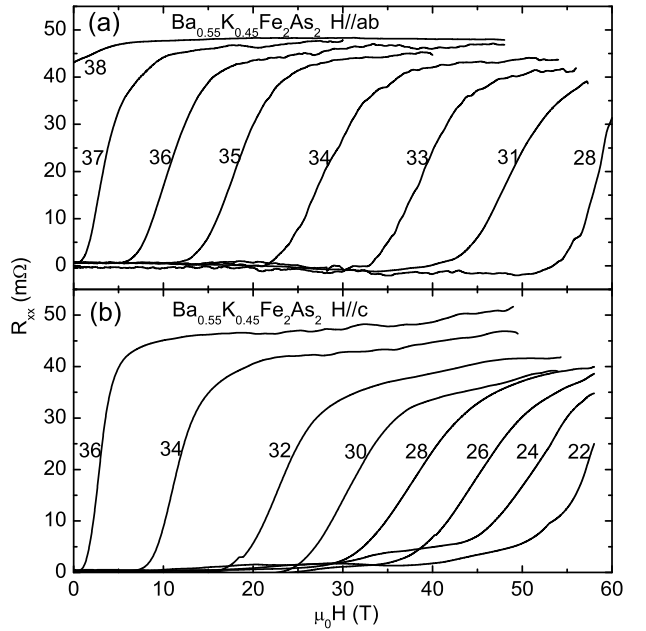}
\caption{Magnetic field dependence of the electrical resistance at
variant temperatures for $\rm Ba_{0.55}K_{0.45}Fe_2As_2$: (a)
$H//ab$ and (b) $H//c$. The temperatures are labeled in the figure
with a unit of Kelvin.} \label{fig3}
\end{figure}

The evolution of superconductivity with magnetic field is shown in
Fig.\ref{fig2} and Fig.\ref{fig3} for $\rm(Sr,Na)Fe_2As_2$(\#N2) and
$\rm Ba_{0.55}K_{0.45}Fe_2As_2$, respectively. In both figures, the
top panel (a) is for magnetic field in the ab-plane and the bottom
one (b) is for field along the c-axis. Obviously, the
superconducting transition is shifted to lower temperature upon
applying a magnetic field. In the case of $\rm(Sr,Na)Fe_2As_2$, the
critical field required to suppress superconductivity is slightly
different for the two magnetic field orientations applied along the
c-axis and the ab-plane, but superconductivity in $\rm
Ba_{0.55}K_{0.45}Fe_2As_2$ is not yet suppressed even by applying a
field of 60T. The much higher critical field in the K-doped material
is likely attributed to its much higher $T_c$.

\begin{figure*}
\centering
\includegraphics[width=6.0in]{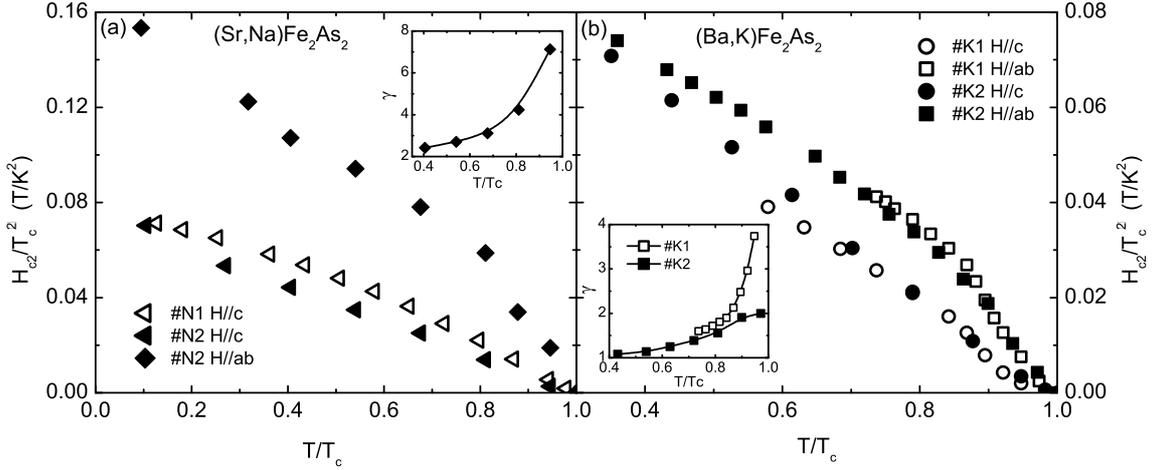}
\caption{The upper critical field of $\rm(Sr,Na)Fe_2As_2$ (\#N1,
\#N2) and $\rm (Ba,K)Fe_2As_2$ (\#K1, \#K2) versus reduced
temperature for $H//c$ and $H//ab$, respectively. The insets show
their anisotropic coefficient ($\gamma$) versus temperature. Note
that the sample \#K1 and \#K2 refer to $\rm
Ba_{0.55}K_{0.45}Fe_2As_2$ measured here and the one reported in
Ref.~\cite{Yuan565}, respectively.} \label{fig4}
\end{figure*}

Fig.\ref{fig4} shows the scaled upper critical field,
$H_{c2}/T_{c}^2$, as a function of reduced temperature, $T/T_c$, for
(a) $\rm(Sr,Na)Fe_2As_2$ and (b) $\rm(Ba,K)Fe_2As_2$, respectively.
Here the value of $H_{c2}$ at each temperature is determined from
the 50\% drop of its normal state resistivity just above $T_c$. The
superconducting transition temperatures ($T_c$), obtained from the
temperature dependence of the electrical resistivity at zero field,
are shown to be 27.6K (\#N1) and 14.8K (\#N2) for
$\rm(Sr,Na)Fe_2As_2$; 36.8K (\#K1) and 28.2K (\#K2) for
$\rm(Ba,K)Fe_2As_2$, respectively. It is noted that the sample \#N1
was unfortunately broken while changing the field orientation to
$H\parallel ab$ and, therefore, only the data with magnetic field
applied along the \emph{c}-axis is shown. The upper critical fields
for sample $\rm Ba_{0.55}K_{0.45}Fe_2As_2$ (named as \#K1 here), as
well as the one reported in Ref~\cite{Yuan565} (sample \#K2), are
included in Fig.~\ref{fig4}(b). From Fig.~\ref{fig4}, one can see
that the upper critical fields of these two compounds show good
scaling behavior with $T_c$, indicating that the large upper
critical field is an intrinsic feature of the iron pnictides. In
this case, the actual upper critical field may strongly depend on
the superconducting transition $T_c$ and, therefore, on the sample
compositions or sample quality. For example, the upper critical
field $H_{c2}(0)$ of $\rm Ba_{0.55}K_{0.45}Fe_2As_2$ (\#K1) is
determined as 130T according to the scaling behavior together with
inputs of $H_{c2}(0)$ = 70T and $T_c = 28$K from sample
\#K2~\cite{Jo}. Such a large $H_{c2}(0)$ far exceeds the
weak-coupling Pauli paramagnetism limit, given by $H_{c2}^p =
1.86T_c =$ 68T, which might indicate an unconventional type of
superconductivity in iron pnictides. Whether the large upper
critical field is attributed to the Pauli paramagnetism~\cite{Jo} or
the orbital effect~\cite{Yuan565} remains unclear. However, the weak
anisotropy of $H_{c2}(T_c)$ seems to support the later as argued in
Ref.~\cite{Yuan565}.

The anisotropic coefficient $\gamma$, defined as
$\gamma=H_{c2}^{H\parallel ab}/H_{c2}^{H\parallel c}$, is plotted in
the insets of Fig.~\ref{fig4}. One can see that, in all these 122
compounds, the value of $\gamma$ decreases monotonically with
decreasing temperature, showing rather weak anisotropy at low
temperatures. Such a weak anisotropy of $H_{c2}(T_c)$ is remarkably
distinct from those shown in the high $T_c$ cuprates and the organic
superconductors~\cite{Vedeneev,organic}. The curvatures of
$H_{c2}(T_c)$ in $\rm(Sr,Na)Fe_2As_2$ and $\rm(Ba,K)Fe_2As_2$ are
somewhat different (see Fig.~\ref{fig4}): in the former one (sample
\#N2), the upper critical field shows linear temperature dependence
for fields applied either in the ab-plane or along the c-axis, but
the two curves of $H_{c2}(T_c)$ merge together at low temperature in
the later compound. Evidence from both experimental
measurements~\cite{Matano,Ding} and band structure
calculations~\cite{Singh} has shown that the iron pnictides are
multi-band systems, consisting of electron pockets and hole pockets.
Doping of holes in the compounds as studied here would modify the
electronic structure and, therefore, change the detailed temperature
dependence of the upper critical field. The slightly distinct
anisotropic behavior of $H_{c2}(T_c)$ in these two compounds might
be attributed to the variant doping levels as what we also observed
in Ba(Fe,Co)$_2$As$_2$~\cite{LJ}. It is noted that a pronounced
upturn curvature is observed near $T_c$ in $\rm
Ba_{0.55}K_{0.45}Fe_2As_2$, which might originate from the
multi-band effect of the system as well.

Based on the results measured here for the hole-doped compounds and
those previously investigated for the electron-doped
compounds~\cite{Baily,Kano}, weak anisotropy of superconductivity
seems to be a general feature of the iron pnictides, in particular
for the 122-system and the 11-compound. The anisotropic properties
are usually determined by the underlying electronic structure, which
was found to be rather two dimensional in cuprates and organic
superconductors~\cite{Vedeneev,organic}. However, our measurements
indicate that the electronic structure of the iron pnictides might
be more like three dimensional~\cite{Yuan565}, which has been
confirmed by recent ARPES measurements ~\cite{ding,Vilmercati}. All
these suggest that the inter-layer coupling would play an important
role in understanding the mechanism of superconductivity.

We would also like to point out that the weak anisotropic and large
upper critical field are unique features of the iron-based
superconductors, which will make them very promising materials for
future applications.

\section{Conclusion}
In summary, we have determined the upper critical fields
$H_{c2}(T_c)$ of $\rm(Sr,Na)Fe_2As_2$ and $\rm
Ba_{0.55}K_{0.45}Fe_2As_2$ by measuring the electrical resistivity
in a pulsed magnetic field. It was found that these compounds show
large values of $H_{c2}(0)$ and relatively weak anisotropy of
superconductivity at low temperature in general. The upper critical
fields obtained from variant samples can be well scaled with its
superconducting transition $T_c$.

\section{Acknowledgements}
This work was supported by NSFC (No. 10934005), the National Basic
Research Program of China (No. 2009CB929104), the PCSIRT of the
Ministry of Education of China, Zhejiang Provincial Natural Science
Foundation of China and the Fundamental Research Funds for the
Central Universities. Work at NHMFL-LANL is performed under the
auspices of the National Science Foundation, Department of Energy
and State of Florida.

\end{document}